# Peculiar Bi-ion dynamics in $Na_{1/2}Bi_{1/2}TiO_3$ from terahertz and microwave dielectric spectroscopy

Jan Petzelt[a]*, Dmitry Nuzhnyy[a], Viktor Bovtun[a], Marek Paściak[a], Stanislav Kamba[a], Robert Dittmer[b], Šarunas Svirskas[c], Juras Banys[c] and Jürgen Rödel[b]

[a]*Institute of Physics ASCR, Na Slovance 2, 182 21 Praha 8, Czech Republic;*
[b]*Institute of Materials Science, Technische Universität Darmstadt, 64287 Darmstadt, Germany;*
[c]*Faculty of Physics, Vilnius University, Sauletekio 9, Vilnius LT-10222, Lithuania*

Dynamics of the main dielectric anomaly in $Na_{1/2}Bi_{1/2}TiO_3$ (NBT) was studied by time-domain THz and microwave spectroscopy, using also previously published data and their new overall fits. Above the dielectric maximum temperature $T_m \approx 600$ K, the response consists of coupled sub-THz oscillator and a relaxation mode, assigned to strongly anharmonic Bi-ion vibrations and hopping, whose slowing down explains the paraelectric-like permittivity increase to $T_m$. Below $T_m$, the main relaxation continues slowing down and additional relaxation, assigned to quasi-Debye losses, appears in the $10^{11}$ Hz range. The oscillator hardens on cooling and takes over the whole oscillator strength. The permittivity decrease below $T_m$ is caused by the reduced strength of the relaxations due to dominance of the rhombohedral phase within the coexistence region with the tetragonal phase. The anharmonic dynamics of Bi is supported by previous structural studies. NBT represents a hybrid between standard and relaxor ferroelectric behaviour.

## 1. Introduction

$Na_{1/2}Bi_{1/2}TiO_3$ (NBT) and its solid solutions with barium titanate $BaTiO_3$ (BT) (abbreviated as NBT-BT) are presently the most studied materials with high piezoelectric response as non-toxic lead-free materials, which can substitute the traditional lead-containing materials like PZT and PMN-PT in piezoelectric applications.[1] NBT is known since the fifties and its basic physical properties published before 2005 were summarized e.g. in the review by Isupov.[2] It is one of the very few studied complex perovskites with two aliovalent ions at the A-site of the $ABO_3$ structure (most of the studied complex perovskites concern the mixed B-site solid solutions). It is ferroelectric at room temperature, but undergoes at least two phase transitions at higher temperatures in the 400–800 K range with a very complex structural and dielectric behaviour, exhibiting antiferroelectric (possibly ferrielectric), relaxor ferroelectric and ferroelastic properties. At still higher temperatures it transforms into a simple cubic perovskite structure with a paraelectric dielectric behaviour.

---

*Corresponding author. Email: petzelt@fzu.cz

Let us briefly describe the structural and dielectric behaviour of NBT starting from the parent high-temperature simple cubic paraelectric phase as it appears to date, since many relevant papers have appeared recently, which, however, did not always consider all the rich literature available. On cooling from ~900 K the low-frequency permittivity (at 1 MHz, because at lower frequencies it is usually influenced by the finite conductivity of the sample [3]) first increases following the Curie-Weiss law (with a rather high Curie constant $C \approx 5 \times 10^5$ K) down to about 800 K [2–6], where it undergoes a phase transition into a weakly polar tetragonal phase with a unit-cell doubling along the tetragonal axis (space group P4bm, $Z = 2$) [7] originating from the instability at the $M$-point of the Brillouin zone ($k = (1/2, 1/2, 0)a^*$) observed in quasielastic neutron scattering.[8] This transition is not clearly seen in the dielectric behaviour, but a smooth deviation from the Curie-Weiss law sets in below ~800 K up to the diffuse maximum of the permittivity near 600 K [356]. There is a weak dielectric dispersion below 1 MHz [3], but at higher frequencies it appears pronouncedly [9]. However, no high-frequency data above ~550 K are available. Close but below the permittivity maximum near ~600 K another phase transition appears with a steep drop down in the permittivity and with an incommensurate modulation near the $\Gamma$-point along the tetragonal axis [10] combined with appearance of a rhombohedral ferroelectric phase. The latter phase originates from instability at the *R*-point of the Brillouin zone (***k*** = (½, ½, ½)***a****) also revealed in quasielastic neutron scattering below this point.[4,5,8,10] But unlike the incommensurate phase, its space group *R*3*c*, *Z* = 2, is not subgroup of the tetragonal phase, being only subgroup of the parent cubic phase. Therefore the tetragonal and rhombohedral phases appear to coexist with each other in a rather broad temperature interval (e.g. ~530–670 K,[7] but the numbers differ in different papers) and with a substantial thermal hysteresis. The region below ~550 K is also specified by an appreciable dielectric dispersion below 1 MHz on the overall decreasing permittivity.[2–3456] On further lowering the temperature, the tetragonal phase gradually disappears, but the modulation remains and saturates with about 6-nm periodicity below ~400 K keeping the modulation axis along the tetragonal axis [001] of the vanishing tetragonal phase and coexisting down to cryogenic temperatures with the rhombohedral phase, which has the polar *c*-axis along the body diagonal direction [111] of the cubic cell.[4–56,8,10] Other authors revealed a rhombohedral polar nanodomain pattern ordered quasiperiodically along the [110] direction [11] or as a periodic coexistence of rhombohedral and orthorhombic [12] or monoclinic *Cc* [13–15] phases along this direction. It can be described also as a periodic coexistence of nanodomains with antiphase $O_6$ octahedra tilting ($a^-a^-c^-$ in the Glazer's notation,[16] representing the rhombohedral phase) with local in-phase $a^-a^-c^+$ tilting, combined with <111> cation displacements correlated along both <111> and <100> directions.[17] However, it is important to note that on poling even the ceramic sample could be transformed into a stable neat rhombohedral phase.[15]

To understand better the high-temperature paraelectric behaviour of NBT, in our earlier paper [9] it was proposed that in the cubic phase the Na and Bi ions are locally ordered in an alternating way along all the crystallographic axes. The local symmetry remains still cubic, but the space group changes from $Pm\bar{3}m$ ($Z=1$) to $Fm\bar{3}m$ with two formula units in the primitive unit cell ($Z_{prim}=2$). In such a case the phase transition from cubic to rhombohedral phase could be considered as an equitranslational transition (ferrodistortive, i.e. without a change of the unit-cell volume), which would easier allow to understand the Curie-Weiss anomaly of the permittivity. Without such an ordering the ferroelectric transition should be considered as a triggered one by a primary order parameter at the $R$-point of the Brillouin zone (antiphase tilting of the neighboring octahedra about all cubic crystallographic axes), which itself cannot account for the pronounced dielectric maximum. Considering the local ordering, still two order parameters of $F_{1u}$ and $A_{2u}$ symmetry are needed to describe the symmetry change,[9] the former one representing the infrared (IR) active phonon eigenvector of predominant A-TiO$_6$ shifts (Last mode of the simple cubic perovskite structure [18]) and the latter one representing the $R$-point oxygen octahedra tilts in the simple cubic phase. The frozen-in ion displacements at room temperature of both the mode eigenvectors with respect to the cubic phase structure are comparable,[7,9] therefore it is difficult to determine which of the parameters is the primary one and which is triggered.

The local A-site ion ordering was first proposed by Siny [19] to explain the Raman spectra in the cubic phase and was also suggested for explaining the modulated phase within the Landau theory.[10] Also the selection rules for understanding the IR response [9,20] support such local ordering. However, most of the structural studies could not reveal any ordering, only in the transmission electron microscopy (TEM) diffraction experiment [21] the ½(*ooo*) superstructure reflections (*o* – odd Miller indices) were detected up to ~900 K, which confirms such ordering. In Ref. [20] it was shown that these superreflections at room temperature remain present even in solid solutions of NBT with $SrTiO_3$, which transforms the structure into a cubic one. Therefore they cannot be explained by $TiO_6$ octahedra tilting solely. It appears that the TEM technique, as well as the IR and Raman spectroscopic techniques, is more sensitive to local structure than the X-ray or neutron diffraction experiments.

The way of possible local ordering of A-ions (chemical order) and its influence on the ferroelectric structure in the ground state (at 0 K) was also studied by first-principles calculations based on density functional theory by Gröting et al.[22,23] At ambient pressure a coexistence of tetragonal and rhombohedral phase appears preferable (in agreement with the experimentally observed incommensurate modulation down to low temperatures) and depends on the underlying chemical order: the ordered regions prefer tetragonal structure and disordered regions the rhombohedral one. But interestingly, among six different local A-ion orderings in a 2x2x2 structural supercell the ordering with two neighbouring cells of pure Bi and pure Na seems to be preferable, as already earlier suggested from high-pressure Raman spectroscopy,[24] rather than the one to one alternation of the cells. However, this picture would require that even the local chemical order is modulated as the structural modulation wave, which appears not to be probable. Namely, the chemical order is expected to be frozen from rather high temperatures (of the order of 1000 K), whereas the modulation appears only below ~600 K.[10]

To answer the question: what is responsible for the main dielectric anomaly near 600 K from the point of view of the lattice dynamics, one has to consider the temperature dependences of optical phonons and their dielectric response. The IR response, analysed from IR reflectivity spectra (10–430 K) and transmission terahertz (THz) spectra (10–300 K), show five phonon modes, of which only the lowest-frequency one, which was overdamped, was softening on heating from ~50 cm$^{-1}$ at 10 K to ~20 cm$^{-1}$ at 430 K.[9] But its contribution to permittivity remained below ~300, much less than the low-frequency permittivity (at 1 MHz), which amounts to ~1500 at 430 K. An overdamped central mode with culminating strength near 650 K and halfwidth comparable to the IR soft-mode frequency was also observed in Raman spectra [25] and assigned to critical scattering due to heterophase fluctuations between the coexisting phases, which should be responsible for the main dielectric anomaly. Similar overdamped phonon excitations were revealed also in the inelastic neutron scattering spectra at the $\Gamma$-point as well as $M$- and $R$-points of the Brillouin zone.[4,5,8] The intensity maximum of the Bragg reflection at the $M$-point coincides with the permittivity maximum ($T_m \approx 600$ K), where the $R$-point Bragg peak and $\Gamma$-point diffuse scattering set in [4]. The dynamics near the $\Gamma$-point is complicated by the coupling with the transverse acoustic (TA) branch, but the overdamped $\Gamma$-point mode appears to soften towards ~670 K.[4] However, in Ref. 4 no correlation is discussed (neither cited) with the available IR and Raman data even if the frequency range is comparable. It appears that the phase transitions in NBT are driven mainly by instabilities at the $M$ and $\Gamma$-points, but the main dielectric dispersion, which accounts for the permittivity maximum, is caused by another relaxation in a still lower frequency range, which was not yet studied. Therefore we decided to measure the time-domain THz spectra of NBT in a broader temperature range and combine them with additional microwave (MW) measurements to get a more complete picture of the dynamics responsible for the main dielectric anomaly.

## 2. Experiment

The bulk ceramic NBT samples were prepared by a standard mixed-oxide route with 3 h sintering at 1150 °C.[9,26,27] Their complex dielectric properties were determined in the high-frequency (HF, taken from Ref. 9), MW and THz range. At HF (1 MHz–1.8 GHz), complex impedance of the rod-shaped cylindrical sample (~5 mm long and 0.5 mm in diameter) with evaporated gold electrodes on its flat surfaces was measured in the temperature range 100–570 K using a computer controlled dielectric spectrometer with a Novocontrol BDS 2100 coaxial sample cell and a HP 4291B impedance analyser. The complex dielectric permittivity was calculated taking into account the electromagnetic field distribution in the sample.[28]

At higher MW frequencies (25–37 GHz) the sample was measured in the waveguide. The scalar reflection $|R|$ and transmission $|T|$ coefficients were determined using the scalar network analyser Elmika R2400. Rod-shaped cylindrical and rectangular samples in the rectangular waveguide with $TE_{10}$ propagating mode were measured in the temperature range 125–520 K. The samples were placed in the middle of the waveguide perpendicularly to its wider wall, where the electric field component has a maximum (magnetic field minimum). Three samples of different sizes were used (~5 mm long and 0.5 mm in diameter, 0.16 x 0.18 mm$^2$ and 0.08 x 0.10 mm$^2$ in cross section) to ensure correct measurements of the broad range of permittivity values. To obtain the dielectric permittivity $\varepsilon'$ and the imaginary part (loss) $\varepsilon''$, a system of the nonlinear equations $|R| = f(\omega, \varepsilon', \varepsilon'')$, $|T| = f(\omega, \varepsilon', \varepsilon'')$ was solved using the modified Newton optimization method.[28] The rectangular samples were approximated as cylinders, which produces small errors (< several %), smaller than the overall accuracy of the MW measurements.

To measure dielectric response in the THz range (190 GHz–1 THz) on both sides polished NBT sample with the thickness of 44 μm, we used a quasioptic time-domain THz transmission technique, based on a Ti:sapphire femtosecond laser, which generates linearly polarized THz probing pulses in a photoconducting switch TeraSED. The detection of the transmitted THz pulses is based on electro-optic sampling using a 1-mm thick plate of the [110] ZnTe crystal.[29] For high-temperature experiment, the commercial high-temperature cell (SPECAC P/N 5850) was used.

To fit the experimental data for the complex dielectric function in the broad frequency range, we used the phenomenological Cole-Cole (CC) relaxations and one damped harmonic oscillator (HO) in the THz range:

$$\varepsilon^*(\omega) = \varepsilon_\infty + \sum_j \frac{\Delta\varepsilon_j}{1+(i\omega/\omega_{Rj})^{1-\alpha_j}} + \frac{\Delta\varepsilon_{HO}\omega_{HO}^2}{\omega_{HO}^2 - \omega^2 + i\gamma_{HO}\omega} \qquad (1)$$

where $\Delta\varepsilon_j$ is the dielectric strength of the $j$-th CC relaxation, $\varepsilon_\infty$ is the higher-frequency contribution to permittivity, $\omega$ is the linear frequency, $\omega_{Rj}$ is the mean relaxation frequency of the $j$-th relaxation, and $\Delta\varepsilon_{HO}, \omega_{HO}$ and $\gamma_{HO}$ denote the dielectric strength, frequency and damping of the oscillator, respectively. Parameter $\alpha_j$ characterizes the distribution of relaxation times around the $j$-th relaxation, $\alpha_j = 0$ corresponds to the Debye relaxation, $\alpha_j \to 1$ corresponds to very broad distribution of Debye relaxations (constant-loss spectrum).

## 3. Results and fitting

Dielectric data evaluated from the THz measurements in the 300–900 K range are provided in Figure 1. One can see that the dielectric loss is monotonically increasing with temperature, whereas the slope in the spectra is changing from a positive one over zero (constant loss spectrum at 400 K) to strongly negative one at high temperatures.

The permittivity spectra show always a negative slope, which monotonically increases with increasing temperature and with decreasing frequency, even if the dependences are not very pronounced. For fitting we divided our results in two temperature ranges: from $T_m \approx 600$ K up to 900 K, where we have only THz data, which we combine with the published low-frequency (which we consider static) data at ~1 MHz [4,5,30] and fit them together, and from 100 K to ~550 K, where we have in addition the complex impedance data in the 1–200 MHz range (higher frequencies using this measurement technique were not found sufficiently trustable) and waveguide MW data at 25–37 GHz. The resulting spectra together with their fits in the high-temperature range are shown in Figure 2. For fitting we have used the simplest possible model with one strong Cole-Cole relaxation (CC1), which softens from ~0.15 THz at 900 K down to 10 GHz at 600 K. It can be well fitted with Debye relaxation ($\alpha = 0$) down to 750 K, but it broadens up to $\alpha \approx 0.2$ at 600 K. To fit the THz data well, an additional heavily damped or overdamped harmonic oscillator (HO) was needed near but below the THz range. Because for the overdamped oscillators the damping constant $\gamma$ and the frequency $\omega_{HO}$ (restoring force constant) are strongly correlated and cannot be experimentally determined accurately, during all the fits the damping was kept constant at $\gamma = 1.9$ THz. Then the temperature dependences are revealed either in $\omega_{HO}$ or in the quantity $\omega_{HO}^2/\gamma$, which corresponds closely to the frequency of its dielectric loss maximum. The latter frequency could be much more accurately determined experimentally. A similar situation appears in inelastic scattering experiments, where the overdamped oscillator appears as a central peak whose half width corresponds to $\omega_{HO}^2/\gamma$ and the $\omega_{HO}$ and $\gamma$ parameters cannot be determined separately reliably, either. The contribution of higher-frequency phonons from the IR data [9] and optical permittivity was demonstrated to be essentially temperature independent and was fixed at $\varepsilon_\infty = 50.5$ for all the temperatures.

We note that a simple fit without any relaxation with only one overdamped oscillator was not satisfactory, because the slope in the fitted spectra in the THz range was steeper than that of experimental spectra.

The lower temperature dielectric responses with their fits are provided in Figure 3. Here two Cole-Cole relaxations (CC1 and CC2) were needed in addition to the sub-THz HO. The additional relaxation CC2 appears in the 100 GHz range. It weakens on cooling without appreciable temperature dependence of its frequency and below room temperature its effect on the spectra becomes negligible. Its Cole-Cole exponent remains below $\alpha = 0.08$, i.e. close to Debye relaxation. The existence of it is enforced by the MW data, which we could measure only up to 520 K. Its presence above this temperature cannot be excluded, our data being insufficient to decide it. The main relaxation CC1 continues the slowing down by about two orders of magnitude, weakens and smears on cooling with $\alpha$ increasing from 0 at 450–550 K to ~0.13 at 100 K. Finally, at 10 K its dielectric contribution is vanishing and the static permittivity (~150) is fully due to the contribution $\Delta\varepsilon_{HO} + \varepsilon_\infty$.[9] Temperature dependences of the characteristic frequencies (peaks of the individual dielectric loss spectra contributions) of the CC1 and CC2 and of the HO (i.e. $\omega_{HO}^2/\gamma$) are depicted in Figure 4 and their dielectric strengths (contributions to the static permittivity) are displayed in Figure 5. It is seen that the HO frequency is not strongly temperature dependent and does not soften towards the dielectric maximum. Its dielectric strength $\Delta\varepsilon_{HO}$ is also roughly constant and stays between 100 and 180. The dielectric maximum at $T_m$ is fully due to the slowing down of the CC1 relaxation, whose frequency $\omega_{CC1}$ reduces by an order of magnitude and its dielectric strength increases by a factor of three. The dielectric strength of the weaker CC2 relaxation below $T_m$ with a constant frequency in the 100 GHz range is much smaller reaching a maximum of ~420 at 520 K. As already mentioned, its behaviour above 520 K remains open for further studies and its presence could be added to the fits, but with ambiguous parameters. However, above ~700 K the response of CC1 overlaps with the possible CC2 and therefore loses any meaning. Moreover, in this high-temperature range the CC1 response overlaps also with the HO response. As seen from Figure 2, in fact only one (slightly asymmetric) loss peak characterizes the overall ε''(ω) spectra in the whole 600–900 K range, but, as already noticed, it was not possible to fit it with only one HO mode.

## 4. Discussion

Let us start the discussion with the behaviour and assignment of the phonon HO mode. At low temperatures below 100 K, where the data are taken from our earlier paper,[9] the THz response is fully due to this mode, which is nearly overdamped with the loss maximum frequency $\omega_{HO}^2/\gamma \approx 40$ cm$^{-1}$ and damping $\gamma \approx 60$ cm$^{-1}$. On increasing temperature it appreciably softens (even if not completely monotonically) down to $\omega_{HO}^2/\gamma \approx 7$ cm$^{-1}$ near 600 K, becoming strongly overdamped ($\gamma \approx 60$ cm$^{-1}$ is taken temperature independent). Its dielectric strength (contribution to the static permittivity) increases from ~100 at 10 K up to ~180 at 600 K only, much less than the low-frequency permittivity values. Above $T_m$, surprisingly, the HO does not change pronouncedly, even if the accuracy of its fitting is reduced by the approaching CC1 relaxation (see Figure 4), which substantially contributes to the THz response at high temperatures. Therefore the scatter in the HO frequencies above 600 K is probably not relevant and, anyway, no appreciable softening towards the dielectric maximum in the paraelectric phase is seen. The dielectric maximum above $T_m$ is therefore solely due to the slowing down (softening) of the CC1 relaxation. The temperature dependence of $\omega_{CC1}$ from 900 to 600 K follows approximately the Arrhenius law, i.e. it could be thought to be thermally activated with $\omega_\infty \approx 24$ THz and activation energy ~0.4 eV. But such activation energy (~4650 K) is too high for a strong thermally activated process. Nevertheless, in the linear frequency scale the slowing down of $\omega_{CC1}$ is almost linear, with the extrapolated softening temperature ~550 K. Such a slowing down is expected for a classical order-disorder ferroelectric transition.[31] However, unlike such phase transitions, below $T_m$ $\omega_{CC1}$ continues slowing down by another two orders of magnitude with two different much smaller activation energies (~0.12 and ~0.024 eV in the 500–300 K and 300–100 K interval, respectively), combined with a dramatic decrease of its dielectric strength. This is more characteristic of thermally activated processes. Let us mention that even below 1 MHz another weaker, presumably intrinsic CC relaxation was observed,[3] also thermally activated, but with higher activation energy (~0.81 eV below $T_m$ and ~0.6 eV above $T_m$).

The behaviour of the CC1 relaxation could be phenomenologically understood assuming rather strong coupling of the HO with the CC1 relaxation, which could be expected particularly at high temperatures where both modes have very close frequencies. The coupling could transfer the softening from the bare HO into the CC1 mode. This phenomenon leads to well-known central mode effects in the vicinity of displacive phase transitions (for ferroelectric transitions in the paraelectric phase it was reviewed in Ref. [31] and later in [32]). In our case, appearance of the central mode (i.e. CC1 relaxation) occurs in a very broad temperature range, at least 600–900 K, and completely prevents our (dressed) HO from softening.

Considering the assignment of the HO mode, in Ref. [9] it was assumed that it might stem from the R-point phonon, which activates in the IR due to the local A-ion ordering. However, its strength is comparable to the main IR phonons from the cubic perovskite structure. Therefore we already have changed the assignment [18] and assigned it to the E-component of the lowest-frequency TO mode of the simple cubic perovskite structure, which in NBT corresponds to so-called Last mode, i.e. to vibrations of the A-ions against the TiO$_6$ octahedra. Because the Bi-ion is much heavier (at. w. ≈ 209) than the Na-ion (at. w. ≈ 23), it is possible to expect a splitting of this mode with the Bi-vibrations having lower frequency. This assignment is in agreement with a recent first-principles calculation by Niranjan *et al*. [6], where all the $\Gamma$-point phonons of the A-ion ordered rhombohedral structure at 0 K were calculated, including their Born effective charges and dielectric strengths. Even if the calculated frequency of this lowest-frequency E(TO1) mode (109 cm$^{-1}$) was appreciably higher and its oscillator strength somewhat lower (5.6·10$^{13}$ Hz compared to the experimental strength of the HO at low temperatures of 1.0·10$^{14}$ Hz, see Figure 6), our HO should be clearly of this Bi – TiO$_6$ octahedra type of vibration perpendicular to the polarization (Last mode of E-symmetry). The mismatch in the frequency may result from the fact that the theoretically derived volume is ~5% smaller than the one observed experimentally, an error typical for local density approximation calculations.

Concerning the assignment of relaxations, the CC1 should be due to the anharmonic vibrations or hopping of some disordered ions. In our case it clearly concerns the Bi-ions, since only this could explain the strong strength transfer from the HO (assigned to Bi-vibrations) to the CC1 above $T_m$. The CC2 relaxation could be assigned to so called quasi-Debye losses.[33,34] It was theoretically predicted for anharmonic non-centrosymmetric structures as due to fluctuation of the thermal-phonon density distribution function and its frequency is given by the mean damping of thermal phonons. We have recently revealed (also from broad-band dielectric fits) a very similar relaxation in Ba(Zr,Ti)O$_3$ relaxors.[35]

The question remains what determines the permittivity decrease below $T_m$. From Figure 5 it is seen that it is due to a dramatic decrease in the strength of the CC1 relaxation below $T_\mathrm{m}$, continuing on further cooling, accompanied by an increase in the HO strength, which can be understood by reducing the coupling between both the excitations. Since it is known that $T_\mathrm{m}$ corresponds to temperature of tetragonal and rhombohedral phase coexistence where on cooling the rhombohedral phase starts to dominate, apparently the coupling between both excitations is preferred in the tetragonal phase. This should be caused by some change of the local strongly anharmonic (probably multi-well) potential for the Bi vibrations and hopping.

The fact that the behaviour of Bi-ions may be a key feature triggering the low-frequency dynamics of NBT is of no surprise. Bi atoms have lone-pair electrons which induce stereochemical activity and make symmetrical atomic environment undesirable.[36] Furthermore, a bond-valence sum (BVS) calculation [37] indicates that Bi-ions are strongly underbonded when occupying positions suggested by the average structure diffraction analyses. In the tetragonal phase [7] BVS for Bi ions is equal to ~2.07 while it slightly rises in the rhombohedral phase (to ~2.3 for the $R3c$ structure reported in Ref. [7]). At the same time, BVS for Na ions remains close to the nominal value of 1. Indeed, large and anisotropic atomic displacement parameters for Bi obtained from structural refinements [38] provide evidence that average structure sites differ substantially from the most probable sites. The fact that Bi cations tend to displace appreciably off-centre in the oxygen polyhedron has been confirmed by the X-ray absorption fine structure (XAFS) experiment.[39] It demonstrates that Bi-O distances as short as 2.22 Å exist in the rhombohedral phase, while the structural model placing Bi cations on the threefold axis predicts the shortest Bi-O distance to be larger than 2.5 Å. A model allowing splitting of Na and Bi positions and comprising multiple sites with partial occupancy for the latter cation has been proven to better explain both the diffraction data and the information from XAFS.[40]

Another evidence of Bi-ion disorder comes from a recent pair distribution function (PDF) analysis.[41] The temperature dependence of A-site cation displacements with respect to the oxygen polyhedra that has been extracted from the neutron total scattering data, allows for an insight into behaviour of Bi-ions. The existence of short Bi-O distances has been confirmed. Interestingly, the distribution of distances has a shoulder at ~2.2 Å for all the measured temperatures (up to 766 K). Displacement directions yielded in the reverse Monte Carlo refinement, change from the <100> tetragonal axes directions (at 766 K) to rhombohedral ones (at 10 K, along the <111> directions) with the most interesting features at intermediate temperatures. The multiple-site and smeared distribution for Bi displacement at 473 K is in agreement with the phase coexistence and with our evidence that the hopping mechanism is still present below $T_m$. Displacement magnitude changes significantly towards smaller values at higher temperatures, but at the same time the distribution of directions gets less well defined. As discussed by Jones and Thomas [42], tetragonal structure of NBT is unique in combining tilts of octahedra about *c* axis and A-site cation displacements along the same axis, as there is no asymmetry in the Bi/Na environment to promote (or respond to) the shift. Bi displacement is then interpreted in terms of the lone pair activity. These facts support our picture of Bi-ions sitting in the anharmonic potential which results in their peculiar dynamical behaviour.

Still another sign of disorder in NBT is manifested in strong diffuse scattering that has been observed in the X-ray diffraction.[13,43] It is interesting to note that in the light of our low-frequency modes assignment and the tendency of Bi-ions to displace from their average positions, a lot of diffuse scattering intensity could possibly be attributed solely to correlations in the shifts of heavy Bi-ions. This idea is yet to be properly explored.

## 5. Conclusions

Our THz and MW measurements with the NBT ceramics enabled us to analyse the unusual features of the dynamics connected with the strong but diffuse maximum of the permittivity near $T_m \approx 600$ K, where no clear phase transition appears. Above $T_m$ it is caused by a pronounced softening of the CC1 relaxation from $\sim 10^{11}$ to $\sim 10^{10}$ Hz, which presumably takes its strength and softening from coupling with an overdamped sub-THz phonon mode playing the role of soft mode. However, the (dressed) mode (HO) does not reveal any softening above $T_m$, but undergoes hardening and strengthening on cooling below $T_m$. Both excitations are assigned to strongly anharmonic dynamics of the Bi ions (split Last mode with the dipole moment perpendicular to the spontaneous polarization, supported by the lattice dynamics calculations in Ref. [6]). The complicated overdamped dynamics of the Bi ions is in agreement with their underbonding and a broad distribution of Bi-O distances. The permittivity decrease below $T_m$ appears due to hardening of the HO and reduced coupling between the HO and the relaxations, presumably connected with the dominance of the rhombohedral phase in the tetragonal-rhombohedral phase coexistence region. The CC1 relaxation is thermally activated below $T_m$, vanishing at low temperatures. The additional CC2 relaxation, which appears in the rhombohedral ferroelectric phase above room temperature in the $10^{11}$ Hz range, is assigned to quasi-Debye losses of anharmonic origin.[33]

The present paper is the first paper devoted to the discussion of the dielectric function and dynamics of NBT in the technically difficult $10^{10}$–$10^{12}$ Hz range. It is clear that NBT does not behave like a classical relaxor ferroelectric, since the usual dielectric dispersion, which provides the shift of permittivity maximum $T_m$ with frequency, sets in only in the GHz range, and the macroscopic symmetry undergoes changes with temperature. But, like in relaxors, the permittivity maximum at $T_m$ is not connected with any phase transition, only with the predominance of the rhombohedral phase in the phase coexistence region.

**Acknowledgements**

The authors thank M. Savinov for the discussion on Arrhenius behaviour. The research was supported by the Czech Science Foundation (project 13-15110S). RD acknowledges support from the German Research Foundation DFG within the collaborative research centre SFB595.

**Figure captions**

Figure 1. Temperature dependence of the complex dielectric function calculated from the THz data.

Figure 2. THz dielectric data, published low-frequency data at 1 MHz [4,5,30] and corresponding fits using Equation (1) in the 600–900 K range at selected temperatures.

Figure 3. HF, MW, and THz dielectric data (including low-temperature data from [9]) and corresponding fits using Equation (1) in the 100–550 K range at selected temperatures. MW data at 125 K and 520 K are shown as data at 100 K and 550 K, respectively.

Figure 4. Temperature dependences of the loss-peak frequencies (relaxation frequencies CC1, CC2 and $\omega_{HO}^2/\gamma$) from the fits in Figures 2 and 3.

Figure 5. Temperature dependences of the dielectric strengths from the fits in Figures 2 and 3.

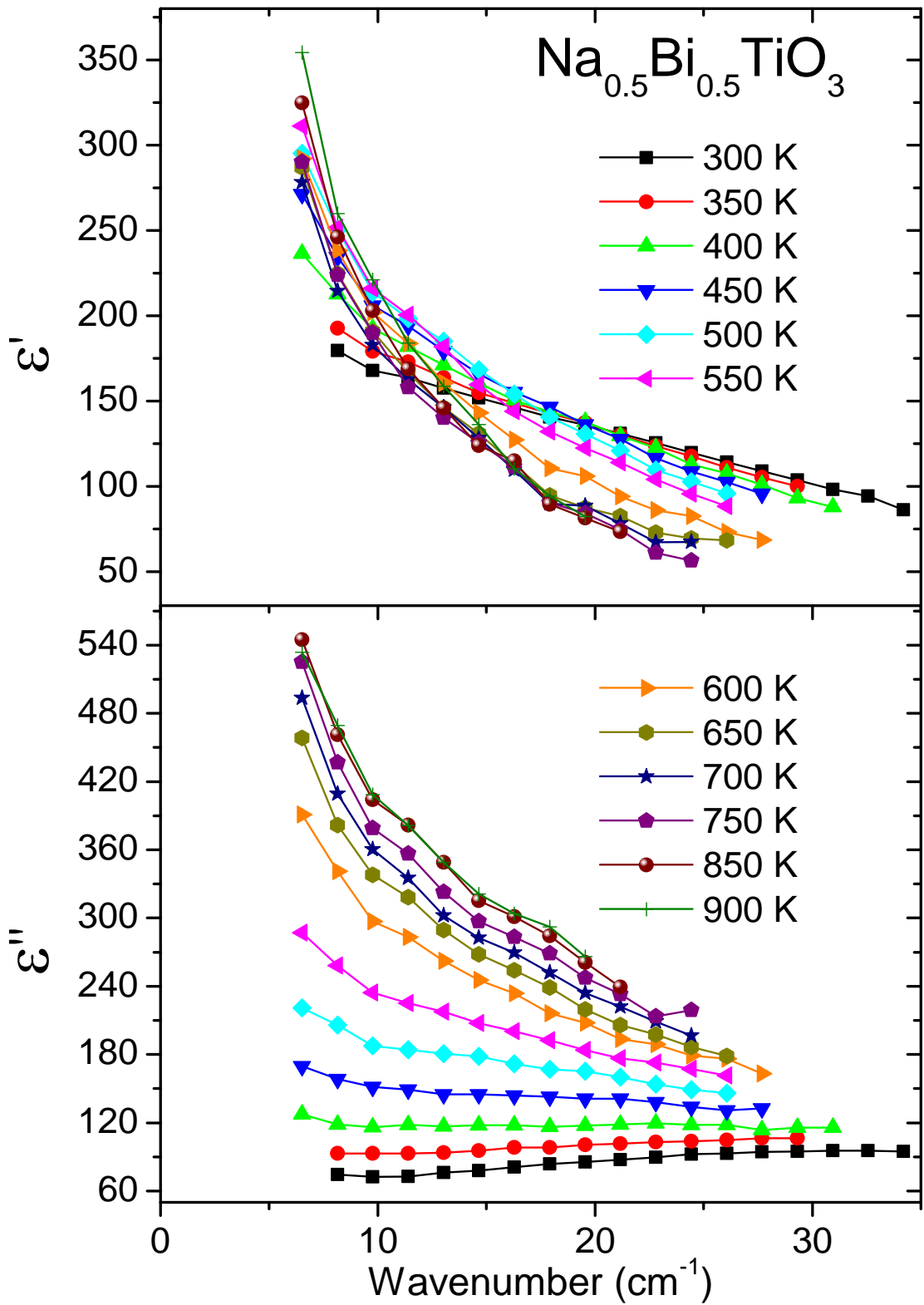

Figure 1. Temperature dependence of the complex dielectric function calculated from the THz data.

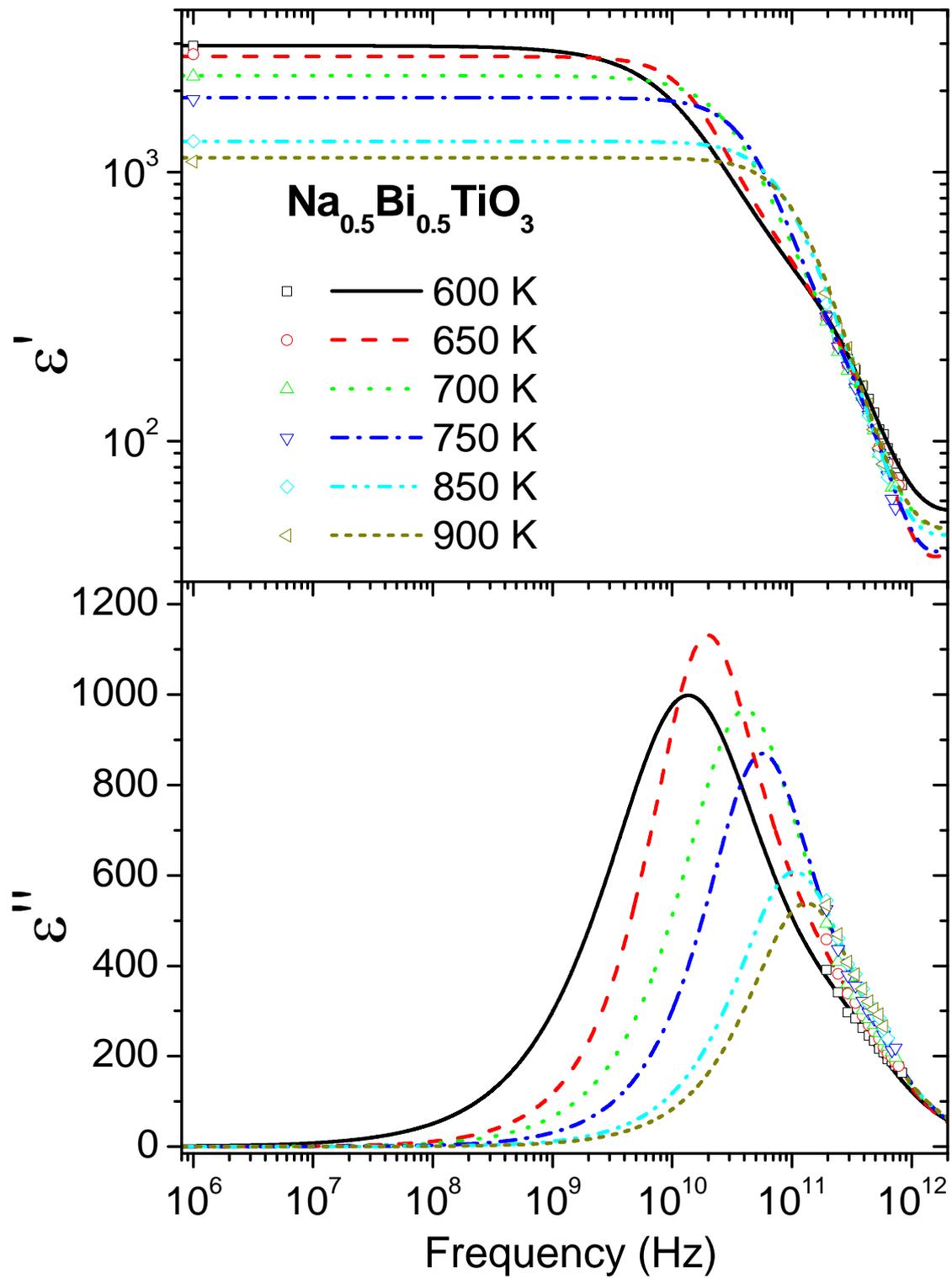

Figure 2. THz dielectric data, published low-frequency data at 1 MHz [4,5,30] and corresponding fits using Equation (1) in the 600–900 K range at selected temperatures.

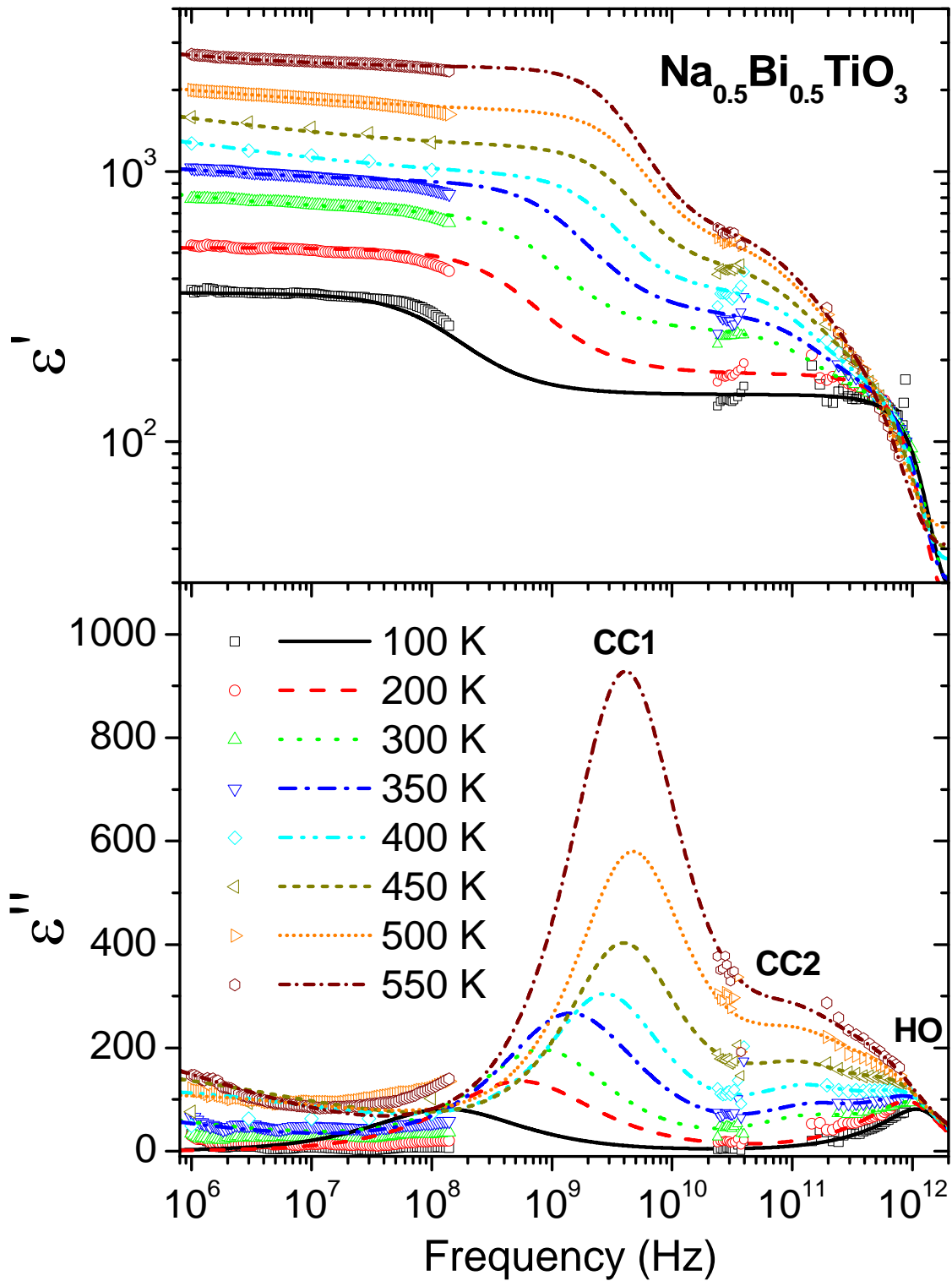

Figure 3. HF, MW, and THz dielectric data (including low-temperature data from [9]) and corresponding fits using Equation (1) in the 100–550 K range at selected temperatures. MW data at 125 K and 520 K are shown as data at 100 K and 550 K, respectively.

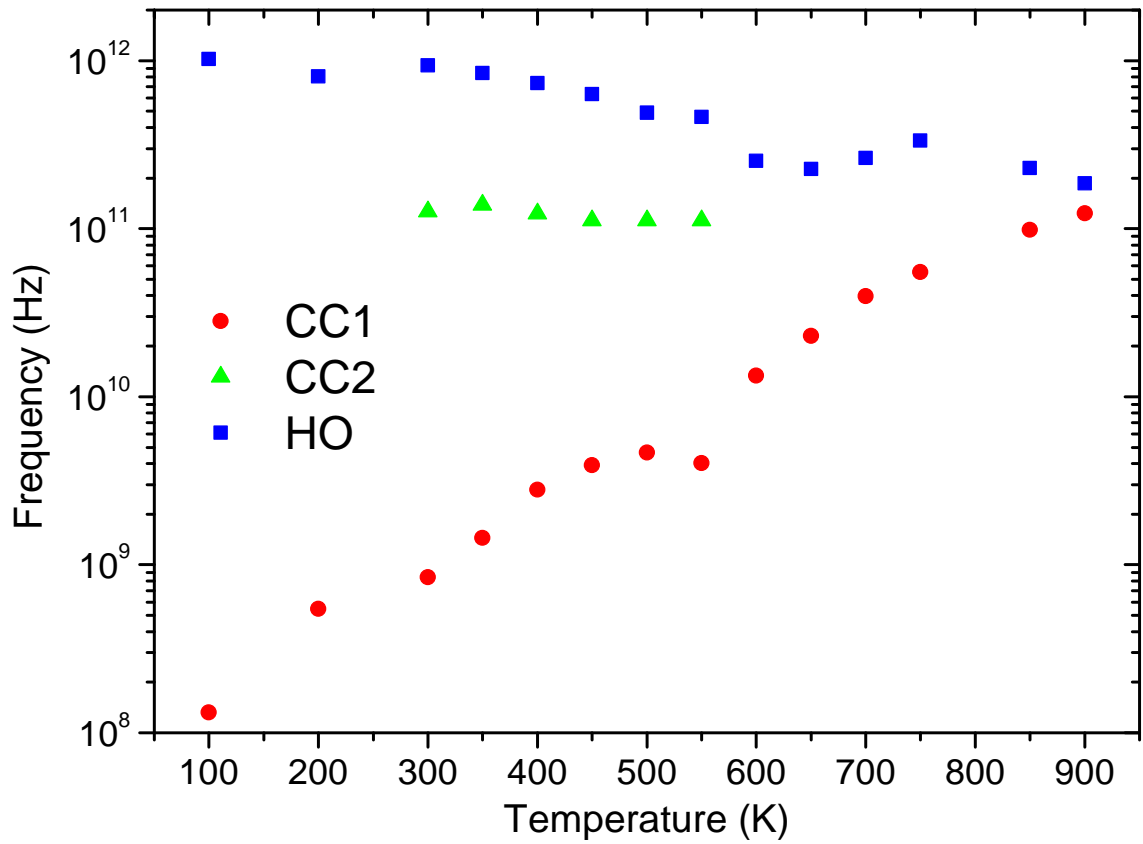

Figure 4. Temperature dependences of the loss-peak frequencies (relaxation frequencies CC1, CC2 and $\omega_{HO}^2/\gamma$) from the fits in Figures 2 and 3.

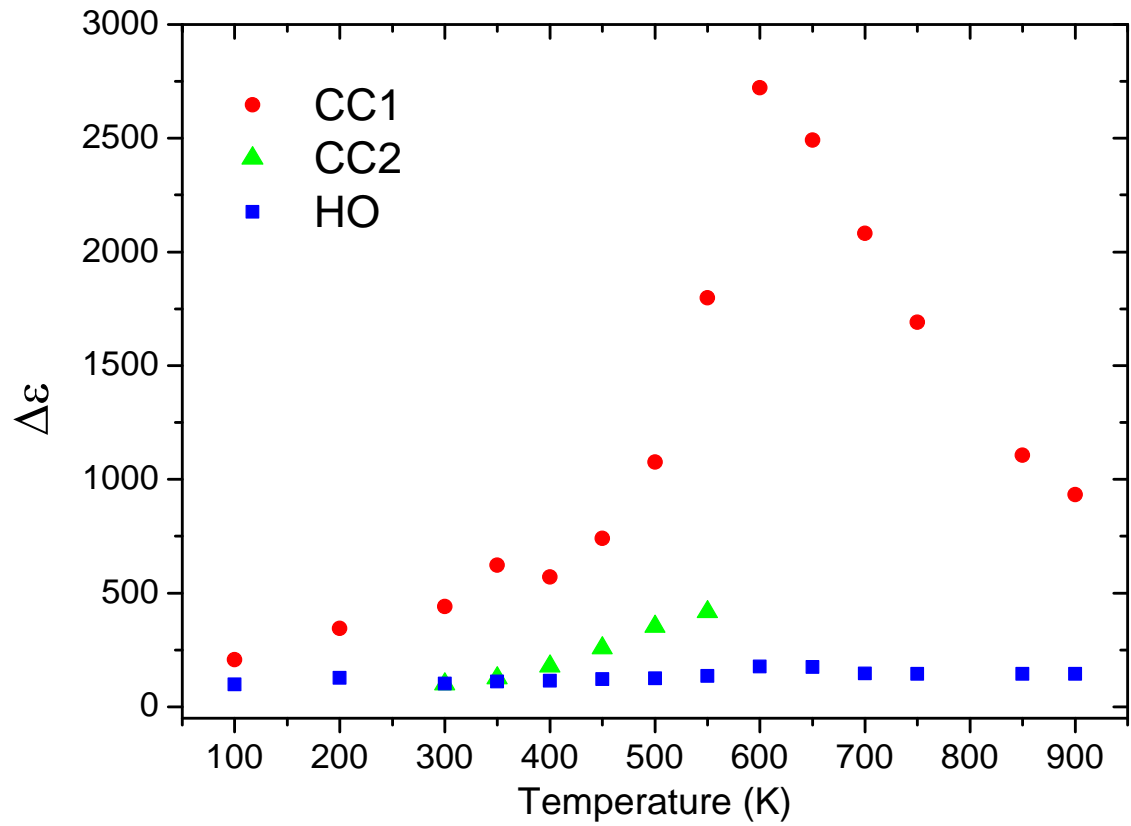

Figure 5. Temperature dependences of the dielectric strengths from the fits in Figures 2 and 3.

**References**

[1]   Rödel J, Jo W, Seifert KTP, Anton E-M, Granzow T, Damjanovic D. Perspective on the development of lead-free piezoceramics. J Am Ceram Soc. 2009;92:1153–1177.

[2]   Isupov VA. Ferroelectric $Na_{0.5}Bi_{0.5}TiO_3$ and $K_{0.5}Bi_{0.5}TiO_3$ perovskites and their solid solutions. Ferroelectrics. 2005;315:123–147.

[3]   Suchanicz J. The low-frequency dielectric relaxation in $Na_{0.5}Bi_{0.5}TiO_3$ ceramics. Mat Sci Eng B. 1998;55:114–118.

[4]   Matsuura M, Ida H, Hirota K, Ohwada K, Noguchi Y, Miyayama M. Damped soft phonons and diffuse scattering in $(Bi_{1/2}Na_{1/2})TiO_3$. Phys Rev B. 2013;87:064109.

[5]   Ge W, Devreugd CP, Phelan D, Zhang Q, Ahart M, Li J, Luo H, Boatner LA, Viehland D, Gehring PM. Lead-free and lead-based $ABO_3$ perovskite relaxors with mixed-valence A-site and B-site disorder: Comparative neutron scattering structural study of $Na_{1/2}Bi_{1/2}TiO_3$ and $Pb(Mg_{1/3}Nb_{2/3})O_3$. Phys Rev B. 2013;88:174115.

[6]   Niranjan MK, Karthik T, Asthana S, Pan J. Theoretical and experimental investigation of Raman modes, ferroelectric and dielectric properties of relaxor $Na_{1/2}Bi_{1/2}TiO_3$. J Appl Phys. 2013;113:194106.

[7]   Jones GO, Thomas PA. Investigation of the structure and phase transitions in the novel A-site substituted distorted perovskite compound $Na_{0.5}Bi_{0.5}TiO_3$. Acta Cryst B. 2002;58:168–178.

[8]   Vakhrushev SB, Isupov VA, Kvyatkovsky BE, Okuneva NM, Pronin IP, Smolensky GA, Syrnikov PP. Phase transitions and soft modes in sodium bismuth titanate, Ferroelectrics. 1985;63,153–160.

[9]   Petzelt J, Kamba S, Fabry J, Noujni D, Porokhonskyy V, Paskin A, Franke I, Roleder K, Suchanicz J, Klein R, Kugel GE. Infrared, Raman and high-frequency dielectric spectroscopy and phase transitions in $Na_{1/2}Bi_{1/2}TiO_3$. J Phys Condens Matter. 2004;16:2719–2731.

[10]  Balagurov AM, Koroleva EYu, Naberezhnov AA, Sakhnenko VP, Savenko BN, Ter-Oganessian NV, Vakhrushev SB. The rhombohedral phase with incommensurate modulation in $Na_{1/2}Bi_{1/2}TiO_3$. Phase Transitions. 2006;79:163–173.

[11] Yao J, Ge W, Yang Y, Luo L, Li J, Viehland D, Bhattacharyya S, Zhang Q, Luo H. Observation of partially incoherent <110> boundaries between polar nanodomains in $Na_{1/2}Bi_{1/2}TiO_3$ single crystals. J Appl Phys. 2010;108:064114.

[12] Dorcet V, Trolliard G, Boullay P. Reinvestigation of phase transitions in $Na_{1/2}Bi_{1/2}TiO_3$ by TEM. Part I: First order rhombohedral to orthorhombic phase transition. Chem Mat. 2008;20:5061–5073.

[13] Kreisel J, Bouvier P, Dkhil B, Thomas PA, Glazer AM, Welberry TR, Chaabane B, Mezouar M. High-pressure x-ray scattering od oxides with a nanoscale local structure: Application to $Na_{1/2}Bi_{1/2}TiO_3$. Phys Rev B. 2003;68:014113.

[14] Gorfman S, Glazer AM, Noguchi Y, Miyayama M, Luo H, Thomas PA. Observation of a low-symmetry phase in $Na_{1/2}Bi_{1/2}TiO_3$ crystals by optical birefringence microscopy. J Appl Cyst. 2012;45:444–452.

[15] Rao BN, Fitch AN, Ranjan R. Ferroelectric-ferroelectric coexistence in $Na_{1/2}Bi_{1/2}TiO_3$. Phys Rev B. 2013;87:060102(R).

[16] Glazer AM. The classification of tilted octahedra in perovskites. Acta Cryst B. 1972;28:3384–3392.

[17] Levin I, Reaney IM. Nano- and mesoscale structure of $Na_{0.5}Bi_{0.5}TiO3$: a TEM perspective. Adv Funct Mat. 2012;22:3445–3452.

[18] Hlinka J, Petzelt J, Kamba S, Noujni D, Ostapchuk T. Infrared dielectric response of relaxor ferroelectrics. Phase Transitions 2006;79:41–78.

[19] Siny IG, Smirnova TA, Kruzina TV. The phase transition dynamics in $Na_{1/2}Bi_{1/2}TiO_3$. Ferroelectrics. 1991;124:207–212.

[20] Lee JK, Hong KS, Kim ChK, Park S-E. Phase transitions and dielectric properties in A-site ions substituted $Na_{1/2}Bi_{1/2}TiO_3$ ceramics (A=Pb and Sr). J Appl Phys. 2002;91:4538–4542.

[21] Trolliard G, Dorcet V. Reinvestigation of phase transitions in $Na_{1/2}Bi_{1/2}TiO_3$ by TEM. Part II: second order orthorhombic to tetragonal phase transition. Chem Mat. 2008;20:5074–5082.

[22] Gröting M, Hayn S, Albe K. Chemical order and local structure of the lead-free relaxor ferroelectric $Na_{1/2}Bi_{1/2}TiO_3$. J Sol State Chem. 2011;184:2041–2046.

[23] Gröting M, Kornev I, Dkhil B, Albe K. Pressure-induced phase transitions and structure of chemically ordered nanoregions in the lead-free relaxor ferroelectric Na$_{1/2}$Bi$_{1/2}$TiO$_3$. Phys Rev B. 2012;86:134118.

[24] Kreisel J, Glazer AM, Bouvier P, Lucazeau G. High-pressure Raman study of a relaxor ferroelectric: The Na$_{0.5}$Bi$_{0.5}$TiO$_3$ perovskite. Phys Rev B. 2001;63:174106.

[25] Siny IG, Husson E, Beny JM, Lushnikov SG, Rogacheva EA, Syrnikov PP. A central peak in light scattering from the relaxor-type ferroelectric Na$_{1/2}$Bi$_{1/2}$TiO$_3$. Physica B. 2001; 293:382–389.

[26] Dittmer R, Jo W, Rödel J, Kalinin S, Balke N. Nanoscale insight into lead-free BNT-BT-xKNN. Adv Funct Mater. 2012;22:4208–4215.

[27] Zhang ST, Kounga AB, Aulbach E, Ehrenberg H, Rödel J. Giant strain in lead-free piezoceramic Bi$_{0.5}$Na$_{0.5}$TiO$_3$-BaTiO$_3$-K$_{0.5}$Na$_{0.5}$NbO$_3$ system. Appl Phys Lett. 2007;91:112906.

[28] Grigas J. Microwave Dielectric Spectroscopy of Ferroelectrics and Related Materials. Amsterdam: Gordon and Breach; 1996. p. 416.

[29] Kužel P, Němec H, Kadlec F, Kadlec C. Gouy shift correction for highly accurate refractive index retrieval in time-domain terahertz spectroscopy. Opt Express. 2010;18:15338–15348.

[30] Yao J, Ge W, Lou L, Li J, Viehland D, Luo H. Hierarchical domains in Na1/2Bi1/2TiO3 single crystals: Ferroelectric phase transformations within the geometrical restrictions of a ferroelastic inheritance. Appl Phys Lett. 2010;96:222905.

[31] Petzelt J, Kozlov GV, Volkov AA. Dielectric spectroscopy of paraelectric soft modes, Ferroelectrics. 1987;73:101–123.

[32] Buixaderas E, Kamba S, Petzelt J. Lattice Dynamics and central-mode phenomena in the dielectric response of ferroelectrics and related materials. Ferroelectrics. 2004;308:131–192.

[33] Gurevich VL, Tagantsev AK. Intrinsic dielectric loss in crystals. Adv Phys. 1991;40:719–767.

[34] Astafiev KF, Tagantsev AK, Setter N. Quasi-Debye microwave loss as an intrinsic limitation of microwave performance of tunable components based on $SrTiO_3$ and $Ba_xSr_{1-x}TiO_3$ ferroelectrics. J Appl Phys. 2005;97:014106–1/8.

[35] Petzelt J, Nuzhnyy D, Savinov M, Bovtun V, Kempa M, Ostapchuk T, Hlinka J, Canu G, Buscaglia V. Broadband dielectric spectroscopy of $Ba(Zr,Ti)O_3$: dynamics of relaxors and diffuse ferroelectrics. Ferroelectrics 2014 in press, arXiv:1312.3131.

[36] Schütz D, Deluca M, Krauss W, Feteira A, Jackson T, Reichmann K. Lone-pair-induced covalency as the cause of temperature- and field-induced instabilities in bismuth sodium titanate. Adv Funct Mat. 2012;22:2285–2294.

[37] Brown ID, Altermatt D. Bond-valence parameters obtained from a systematic analysis of the inorganic crystal structure database. Acta Cryst B. 1985;41:244–247.

[38] Jones GO, Kreisel J, Thomas PA. A structural study of the $(Na_{1-x}K_x)_{0.5}Bi_{0.5}TiO_3$ perovskite series as a function of substitution (x) and temperature. Powder Diffraction. 2012;17:301–319.

[39] Shuvaeva VA, Zekria D, Glazer AM, Jiang Q, Weber SM, Bhattacharya P, Thomas PA. Local structure of the lead-free relaxor ferroelectric $(K_xNa_{1-x})_{0.5}Bi_{0.5}TiO_3$. Phys Rev B. 2005;71:174114.

[40] Thomas PA, Kreisel J, Glazer AM, Bouvier P, Jiang Q, Smith R. The high-pressure structural phase transitions of sodium bismuth titanate. Z Kristallogr. 2005;220:717–725.

[41] Keeble DS, Barney ER, Keen DA, Tucker MG, Kreisel J, Thomas PA. Bifurcated polarization rotation in bismuth-based piezoelectrics. Adv Funct Mat. 2013;23:185–190.

[42] Jones GO, Thomas PA. The tetragonal phase of $Na_{0.5}Bi_{0.5}TiO_3$ – a new variant of the perovskite structure. Acta Cryst B. 2000;56:426–430.

[43] Welberry TR. Diffuse X-ray Scattering and Models of Disorder. IUCr Monographs on Crystallography. Oxford: Oxford University Press; 2004.